# Nonreciprocal transmission of electromagnetic waves with nonlinear active plasmonic metasurfaces

Tianjing Guo[1,#] and Christos Argyropoulos[2,*]

[1]*Institute of Space Science and Technology, Nanchang University, Nanchang 330031, China*

[2]*Department of Electrical and Computer Engineering, University of Nebraska-Lincoln, Lincoln, Nebraska 68588, USA*

[#]*tianjing@ncu.edu.cn*

[*]*christos.argyropoulos@unl.edu*

Nonreciprocity is important for optical information processing, full-duplex communications, and protection of sensitive laser equipment from unwanted reflections. However, it is very challenging to obtain strong nonreciprocal response in optical frequencies, especially when nanoscale configurations are considered. In this work, we solve this problem by demonstrating a compact bifacial plasmonic metasurface that acts as a new ultrathin nonreciprocal transmission filter at near-infrared frequencies. The proposed nanostructure is simple to be practically implemented, since it is composed of two silver nanostripes with different dimensions placed on both sides of an ultrathin active dielectric spacing layer. The introduced gain leads to an exceptional point formation in the linear operation regime, where unidirectional perfect transmission due to loss compensation is achieved from the presented non-Hermitian parity-time symmetric nanoscale system. The demonstrated plasmonic metasurface breaks the Lorentz reciprocity law at the nanoscale when its nonlinear response is considered, mainly due to its spatially asymmetric geometry combined with enhanced nonlinearity. Strong and broadband nonreciprocity with unity transmission contrast from opposite directions is realized under relatively low input laser intensity values. Furthermore, asymmetric nonlinear third harmonic generation with a pronounced contrast is also illustrated by the same nonreciprocal metasurface design. The findings of this work can lead to the realization of various new self-induced nonreciprocal nanophotonic configurations, such as integrated ultrafast switches, ultrathin protective coatings, and asymmetric directional imaging devices.

**Keywords:** nonlinear optics, nonreciprocity, plasmonics, metasurfaces.

## I.    Introduction

The breaking of reciprocity is imperative to several practical photonic applications. As an example, unwanted back reflections can hinder the efficient laser operation [1–4]. Nonreciprocal transmission filters [1] can alleviate this problem by imposing a particular requirement to block light in one direction while allowing light to propagate unperturbed in the opposite direction.

Another widely used nonreciprocal devices are circulators [5], which are important components mainly in microwave systems [6] due to their ability to separate transmitted and received signals. They can control the electromagnetic wave propagation and effectively route it along a preferential direction. Several mechanisms that lead to nonreciprocity have been demonstrated in the literature based on the Faraday effect in magneto-optical media [7–14], optical nonlinear materials [15–23], and structures with time-dependent modulation of refractive index [2,24–28]. Applying a magnetic field to a magneto-optical material is the most common way to break reciprocity, leading to the materials' permittivity becoming an asymmetric anisotropic tensor [29]. However, the requirement of high external magnetic field values, mainly achieved by bulky magnets, and the inevitable loss inherent to magneto-optical materials are significant limiting factors, especially at optical frequencies, making this method impractical to realize high-efficiency nanophotonic nonreciprocal structures. Another representative class of emerging nonreciprocal devices is based on dynamic modulation [2,24,26–28] in a time-dependent system whose operation does not depend on the strength of the bias signal. This approach can relax the demand of special magneto-optical materials, but it still requires sufficiently fast temporal modulation speed that has practical challenges in its efficient realization, especially at higher optical frequencies.

Self-induced nonreciprocity by using the materials' nonlinear response has recently attracted intense interest as an alternative more practical nonreciprocal mechanism in optics, since it does not require an external bias signal. It is well known that nonlinearity can break the Lorentz reciprocity theorem [15,30] but the operation of the resulting device strongly depends on the usually high input intensity values. In addition, this type of device works only for pulsed incident signals [31,32] due to the dynamic reciprocity limitation which means that it cannot operate as perfect isolator when it is simultaneously excited from both sides [31]. Despite these limitations, it has the advantage of not requiring magneto-optical materials or complicated time-modulation schemes, making it an attractive route towards achieving self-induced nonreciprocity. On a relevant note, lately, plasmonic metasurfaces [33–38] have been demonstrated to enhance various optical nonlinear processes [39]. They usually constitute of metallic or dielectric nanoresonators that are specifically arranged in an ultrathin periodic configuration to achieve a desired electromagnetic functionality. Interestingly, it has also been demonstrated that the breaking of Lorentz reciprocity and the resulted nonreciprocal transmission of electromagnetic waves can occur by connecting the nonlinear Kerr effect with parity-time (PT) symmetric active systems, where the formation of exceptional points (EPs) leads to unidirectional perfect transmission due to loss compensation [17,40–42]. A unique characteristic of EPs is that the reflection becomes highly asymmetric by illuminating from opposite directions, while the transmittance is always equal to unity from both directions in the linear regime [43–45]. Nonlinear plasmonic metasurfaces operating near EPs are expected to exhibit enhanced nonreciprocal response due to the strong and asymmetric field enhancement inside their nanostructures [46]. As a relevant example, passive Fano resonators [47,48] based on plasmonic metasurfaces have the ability to boost the interaction between the incident electromagnetic wave and nonlinear material due to their narrow frequency response combined with enhanced electric field values [39]. By properly varying the input intensity, the resonance frequency is shifted because of the system's inherent third-order Kerr nonlinear response. This frequency shift is different under opposite illumination directions in case the Fano resonator geometry is asymmetric [49,50]. By appropriate tuning the incident laser input intensity and operating close to its resonance frequency, it is possible for the transmission to have a minimum value for illumination from one side and maximum value from the opposite direction at the same frequency, leading to a large self-induced nonreciprocal response excited by moderate input intensities [51,52]. However, this nonreciprocal response is usually obtained from relatively large structures (photonic crystals or ring resonators) with several wavelength dimensions, or the nonreciprocal transmission contrast is relatively weak, especially in the case of passive subwavelength configurations [27,49,50,52,53].

In this work, we move one step beyond the limitations of the current designs and propose a new active and reconfigurable compact bifacial plasmonic metasurface to achieve nonreciprocal light transmission operating at near-infrared (IR) frequencies that is triggered by nonlinear self-action due to the Kerr effect. The proposed simple to practically implement metasurface consists of two arrays composed of different dimensions silver nanostripes placed on opposite sides of an ultrathin nonlinear dielectric spacer layer with relatively weak active (gain) response. The gain material not only compensates the unavoidable plasmonic Ohmic losses of metals, but also leads to the formation of a new PT-symmetric non-Hermitian nanophotonic system due to the geometric asymmetry of the structure. The resulted EP formation due to PT-symmetry is utilized to achieve efficient self-induced nonreciprocity in the nonlinear operation regime. The presented system has a Fano resonant response with linear transmission abruptly varying from zero to one over a narrow frequency range around the resonance frequency. This feature can enable high transmission contrast nonreciprocity for opposite direction illuminations by using moderate laser input intensity values. The effect is stemming from the enhanced and asymmetric electric field distribution achieved in the dielectric layer sandwiched between the nanostripes, where the Kerr nonlinearity of the gap material is efficiently triggered. In addition, we find that the nonreciprocal response of the system can be made tunable by varying the input intensity values. Furthermore, we analyze the dynamic reciprocity inherent limitation of the presented nonlinear device by considering a counter propagating additional signal, acting as small perturbation or noise to the system, in conjunction with the main large input signal that triggers the nonreciprocal response. Lastly, enhanced asymmetric nonlinear harmonic generation is demonstrated based on the proposed metasurface by calculating the third harmonic intensity generated under opposite excitation directions. The compact nonreciprocal plasmonic metasurface design presented in our work can find a wide range of applications towards realizing efficient self-induced nonreciprocal nanophotonic systems, including integrated all-optical ultrafast switches, ultrathin protective coatings for lasers and other sensitive equipment, circulators, asymmetric directional image generation, and nonreciprocal transmission filters.

## II.     Geometry and theoretical model

The presented bifacial plasmonic metasurface exhibiting nonreciprocal behavior is based on two periodic arrays of asymmetric silver nanostripes separated by an ultrathin dielectric spacer layer, as shown in Fig. 1. This periodic design can achieve very strong local electric field inside the formed nanogap when excited by much lower input intensities compared to other previous related works [19,49,52]. The bifacial plasmonic metasurface breaks reciprocity mainly due to the spatially asymmetric permittivity values in the nanogap induced by the different nonlinear response triggered from opposite direction illumination. Note that bifacial plasmonic and dielectric metasurface designs have been experimentally verified in the recent literature [46,54–57] for other applications and the currently proposed structure can be realized by using similar fabrication techniques. Here, we optimize and investigate the proposed design by using simulations based on the finite element method [58] that accurately compute its linear and nonlinear response in the case of upward or downward illuminations with propagation directions shown in Fig. 1.

We employ two-dimensional (2D) simulations of the presented periodic nanostructure with unit-cell shown in the inset of Fig. 1. This approximation is valid since the length of each silver nanostripe is much larger than its width. The 2D system can greatly accelerate the simulation process compared to computationally intensive three-dimensional (3D) simulations. The simulation domain is terminated by periodic boundary conditions (PBC) in the x-direction with an optimized period of $p = 250$ nm. Thus, there is only one unit-cell in the simulation domain, consisting of two silver nanostripes on the top and bottom of a nonlinear and active dielectric spacer layer with widths equal to $a = 20$ nm and $b = 160$ nm, respectively. It is worth emphasizing that the two silver nanostripes have different sizes, which satisfy the spatial asymmetry requirement [49,50] to realize nonreciprocal transmission in the nonlinear regime when the proposed device is illuminated from different directions. In order to avoid numerical artifacts and to obtain results closer to a potential fabricated structure, the corners of the nanostripes are slightly rounded during our simulations by using a fillet radius value of 2nm. The ports are placed above and below the unit-cell to create the incident plane waves. In the current work, the proposed device is always illuminated by x-polarized electric field incident plane waves, usually called transverse-magnetic (TM) linear polarization, at normal incidence angle, since the resonance in this structure can be induced only by this polarization [34]. However, the proposed design can become polarization independent, if the nanostripes are replaced by their 3D counterparts, an array of nanocube resonators [59–62] with different sizes located along the upper and bottom sides of the dielectric spacer layer.

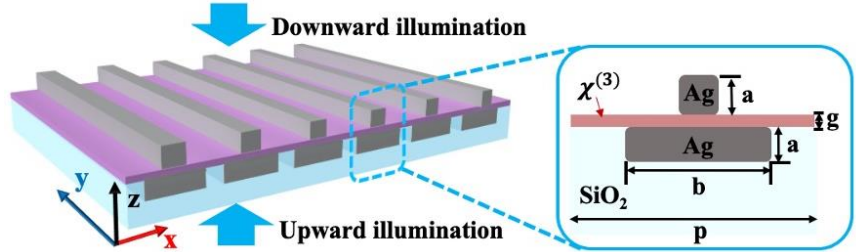

**Figure 1.** Schematic of the proposed self-induced nonreciprocal bifacial plasmonic metasurface.

The linear permittivity of silver follows the Drude model in our simulations: $\varepsilon_{Ag} = \varepsilon_\infty - f_p^2 / f(f - i\gamma)$ that is fitted to experimental data when $f_p = 2175 THz$, $\gamma = 4.35 THz$, and $\varepsilon_\infty = 5$ [63,64]. Note that the geometrical parameters of the proposed metasurface were optimized to make it resonate at the near-IR frequency regime, where the Drude model can be safely applied since it leads to permittivity values very close to the experimentally obtained data [63]. The dielectric spacing layer is composed of an ultrathin nonlinear material ($g = 2$ nm) with a linear relative permittivity of $\varepsilon_L = 5.76$, similar to arsenic trisulfide (As₂S₃). The gain response of this chalcogenide material (As₂S₃) is also introduced to overcome the limitations of the plasmonic system, such as the significant loss of silver stripes, and to achieve high transmittance in our simulations. Gain and nonlinear response can be achieved by this material if it is doped with Erbium or by stimulated Brillouin scattering [65–68]. Importantly, we select to use this chalcogenide material ($As_2S_3$) because it can be grown in thin films [69,70]. Thus, the linear permittivity of the dielectric layer is expressed as $\varepsilon_L = 5.76 + i\delta$, where $\delta = 0.09$ is the used relatively small gain value. Interestingly, several alternative materials or configurations with strong gain coefficients have recently been proposed to operate in the microwave, mid-IR, and visible frequencies [71–74]. In addition, we have verified that the presented active metasurface has a stable response by checking that the pole of this active system resides in the upper half-plane of the imaginary part of complex frequencies [75,76] (not shown here).

In order to achieve a strong nonreciprocal effect based on the presented configuration, we take into account its inherent Kerr nonlinearity. In this case, the permittivity of the dielectric spacing layer depends on the local intensity of the electric field as: $\varepsilon_r = \varepsilon_L + \chi^{(3)} |E_{loc}|^2$, where $\chi^{(3)}$ is the Kerr third-order nonlinear susceptibility, and $E_{loc}$ is the induced local electric field mainly enhanced in the nanogap. Here, we assume that the dielectric spacer layer has a relatively low Kerr nonlinear susceptibility $\chi^{(3)} = 6 \times 10^{-20} m^2 V^{-2}$, which is a typical value for $As_2S_3$ semiconductors [77]. Note that silver also has a nonlinear susceptibility with value $\chi_{Ag}^{(3)} = 2.8 \times 10^{-19} m^2 V^{-2}$ [78] at this frequency range but the fields are weakly penetrating and interacting with silver at near-IR frequencies. As a result, the metallic material has negligible effect on the total nonlinear response of the proposed plasmonic metasurface and is not considered in our simulations. More details about the effect of silver nonlinearity are presented in the Supplementary Material [79]. To propose a more practical design and be able to verify the results in future experiments, the bottom part of the presented nanostructure is embedded inside a semi-infinite dielectric material with relative permittivity of 2.2, similar to silica or quartz, acting as the entire structure's substrate. The proposed configuration can be potentially fabricated by using two step electron beam lithography (EBL) combined with the metal lift-off process on the silica substrate [46].

## III.    Results and discussion

The spectral dependence of the computed transmittance of the proposed metasurface under normal incidence illumination in the linear and nonlinear regimes is shown in Fig. 2(a). The blue line demonstrates the linear response of the structure, which exhibits an ultrasharp Fano resonance response, providing a rapid transmittance value change from zero to around one over a narrow frequency range independent of the illumination direction. The transmittance is slightly larger than one around the resonance frequency of 353.5 THz because the dielectric spacer layer consists of an active material with low gain. In the linear regime, the metasurface is reciprocal and the transmission is always the same under opposite direction illumination. When we take into account the nonlinear susceptibility of the dielectric layer, the increase in the laser input intensity value results in a frequency shift due to the overall nonlinear response of the structure [49]. More importantly, the frequency shift is different from opposite illumination directions since the geometry of our proposed structure is asymmetric. As a result, we obtain high nonreciprocal transmittance contrast under the same frequency range but opposite direction excitations when identical input intensity values are used. The ultrasharp Fano resonant response shown in Fig. 2(a) exists for both the linear and nonlinear cases of the current active metasurface design. It is due to interference of a broad resonance obtained from the periodic bottom nanostripe geometry, when there is no gap, and the narrowband localized gap-plasmon resonance that happens when only the periodic upper nanostripes are used combined with the active gap layer that is placed over a continuous ultrathin (20 nm thick) metallic substrate. Note that realistic frequency dispersion of the gain medium that obeys the Kramers-Kronig relations will also lead to an ultrasharp Fano resonant response as presented in Fig. 2(a).

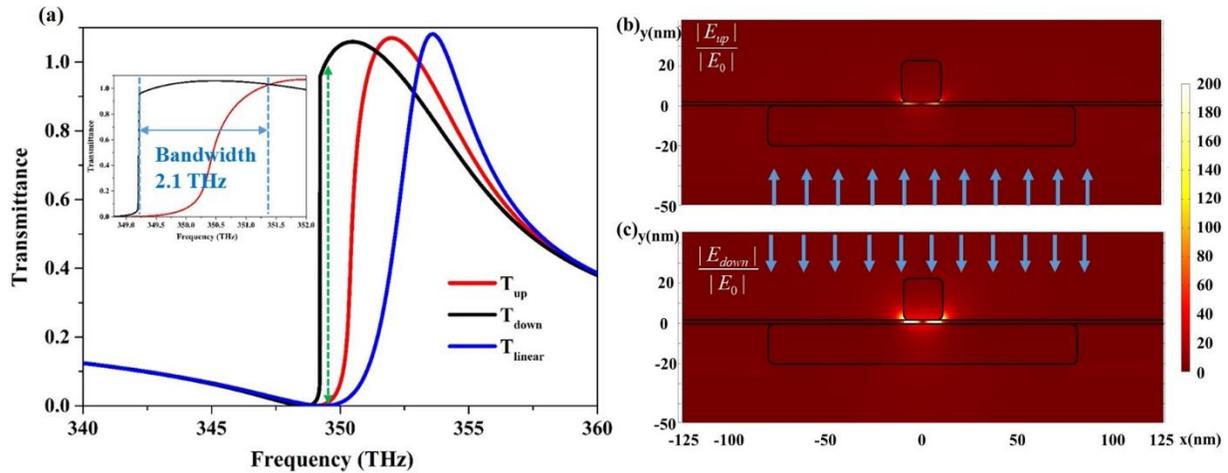

**Figure 2.** (a) Linear and nonlinear responses under upward and downward laser illumination. The zoomed-in plot at the inset demonstrates the bandwidth of the calculated nonreciprocal transmittance. (b)-(c) The asymmetrically induced electric field enhancement distribution at the maximum transmission contrast frequency (green dashed line in caption (a)) in the nonlinear regime under upward and downward illuminations.

To illustrate this interesting nonreciprocal effect, we excite the proposed structure with a normal incident wave propagating either from the upward or downward directions and calculate its transmittance with results shown in Fig. 2(a) by using red (up) and black (down) lines, respectively. It can be clearly seen that very high nonreciprocal transmittance contrast is obtained around 349.4 THz by using relatively low input intensity values equal to 8 MW/cm². For the upward excitation, the transmittance is almost zero at 349.4 THz, while it can reach to one for the downward excitation, which is demonstrated by the green dashed line in Fig. 2(a). The obtained almost perfect nonreciprocal transmission contrast extends over a relative broad bandwidth of 2.1 THz, as shown in the inset of Fig. 2(a) that depicts a zoomed-in plot at the frequencies of strong nonreciprocal transmission. This is an important advantage of the current design compared to relevant previous works [19–21], where the bandwidth was more limited to 0.02 THz [20] and 0.6 THz [21], respectively, especially considering that this type of self-induced nonreciprocity can be achieved only for pulsed illumination. The nonreciprocal effect is further verified by plotting the electric field enhancement distribution, computed in the nonlinear operation, that is asymmetrically induced under different side illuminations with results presented in Figs. 2(b) and (c), respectively. We used an entire unit cell of the bifacial metasurface to demonstrate this result where the strong asymmetric field enhancement is found to be dominant in the nanogap. Notable the field enhancement value in the dielectric spacer region can reach values as high as 200 only in the case of downward illumination. Hence, different field values are induced when the designed structure is excited from opposite directions, meaning that the induced at the nonlinear operation Kerr effect will lead to different transmission response depending on the illumination direction. The resulted asymmetric field distribution and Kerr nonlinear effect triggered by low input intensity values are key factors to obtain the efficient self-induced nonreciprocal transmission based on the presented ultrathin compact plasmonic design.

The presented active plasmonic metasurface system is also interesting in the linear operation, since it consists a new non-Hermitian nanophotonic configuration exhibiting PT-symmetry that can lead to the formation of EPs. To prove this point, we compute the eigenvalues and eigenvectors of the proposed system by deriving the transfer matrix M from the computed scattering matrix S using the relationship [17,75,80]:

$$M = \begin{bmatrix} M_{11} & M_{12} \\ M_{21} & M_{22} \end{bmatrix} = \begin{bmatrix} S_{21} - \dfrac{S_{22}S_{11}}{S_{12}} & \dfrac{S_{22}}{S_{12}} \\ -\dfrac{S_{11}}{S_{12}} & \dfrac{1}{S_{12}} \end{bmatrix}, \tag{1}$$

where $S_{11}$ and $S_{22}$ are the reflection coefficients for the upward and downward directions, respectively, and the transmission coefficients are equal $S_{12} = S_{21}$ in the linear regime due to reciprocity. We prefer to use the transfer matrix method to perform our analysis instead of the scattering matrix formalism [81,82], since it is ideal approach to analyze the currently studied two-port network and study in our future work cascaded bifacial metasurfaces that are expected to achieve even broader and stronger nonreciprocity with lower gain values. The two eigenvalues of the matrix M are calculated by the formula [17,75,83]:

$$\eta_{1,2} = \frac{M_{11} + M_{22}}{2} \pm \sqrt{(\frac{M_{11} + M_{22}}{2})^2 - 1} \ . \tag{2}$$

Different from the orthogonal eigenvectors of a Hermitian system, the eigenvectors of a non-Hermitian system are biorthogonal and are given by the relation [83]:

$$\left\langle \varphi_1^l \mid \varphi_2^r \right\rangle = \left\langle \varphi_2^l \mid \varphi_1^r \right\rangle = 0 \,, \tag{3}$$

where the eigenvectors are calculated by the transfer matrix M following the relation:

$$\langle\varphi_{1,2}^{l}\big| = \begin{pmatrix} \dfrac{\eta_{1,2}-M_{22}}{M_{12}} & 1 \end{pmatrix}, \quad \langle\varphi_{1,2}^{r}\big| = \begin{pmatrix} \dfrac{\eta_{1,2}-M_{22}}{M_{21}} \\ 1 \end{pmatrix}. \tag{4}$$

The evolution of the eigenvalues is computed by Eq. (2) in the complex frequency domain as a function of a small variation in the dielectric material gain coefficient $\delta$ at the linear plasmonic resonance frequency 353.5 THz and is shown in Fig. 3(a). We observe a clear EP formation degeneracy for a gain coefficient of $\delta = 0.09$, which is manifested by the collapse of the two eigenvalues into one point along the unit circle only for this particular gain coefficient value. For comparison, we also present the evolution of the two eigenvalues as a function of the gain coefficient at 354 THz, i.e., slightly off the resonance frequency, with results shown in Fig. 3(b). Interestingly, the two eigenvalues are always separated in this case, demonstrating no EP formation at other frequencies, except at the resonance frequency of the asymmetric plasmonic metasurface. To further verify the EP formation, the absolute values of the two eigenvalues are computed by using Eq. (2) as a function of the incident frequency for a fixed gain coefficient value $\delta = 0.09$ and the results are presented in Fig. 3(c). We find a clear bifurcation point at 353.5 THz, which coincides with the resonance frequency of the linear non-Hermitian system. This is an additional clear indication of an EP formation, which is further verified by computing with Eq. (4) the associated eigenvectors as a function of incident frequency, as depicted in Fig. 3(d). It is found that the eigenvectors again coalesce at the resonance frequency (353.5 THz) for a fixed gain coefficient value 0.09, consisting another unambiguous indication of an EP formation. The strong and asymmetric field enhancement near EPs leads to self-induced broadband and giant nonreciprocal response, however, without the usual need of applying an external magnetic bias in elongated bulky configurations [84].

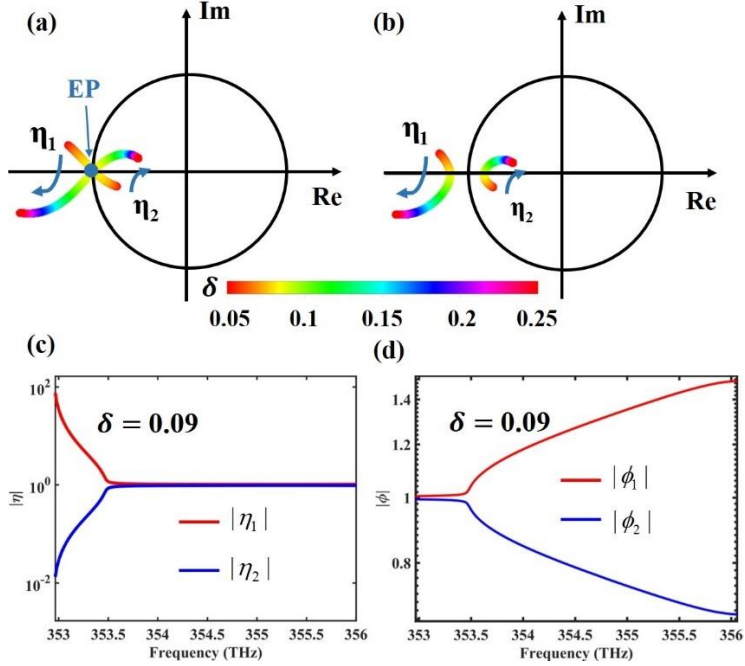

**Figure 3.** (a), (b) Evolution of the complex eigenvalues $\eta_1$ and $\eta_2$ as a function of a small variation in the gain coefficient $\delta$ at (a) the linear plasmonic resonance frequency (f = 353.5 THz) and (b) slightly off this resonance frequency (f = 354 THz). (c), (d) Absolute values of the computed two (c) eigenvalues and (d) eigenvectors as a function of the incident frequency when the gain coefficient is fixed to $\delta = 0.09$.

The linear EP formation frequency point is expected to be detuned and shifted by the Kerr third-order nonlinear effect that is mainly triggered in the dielectric spacer layer where the electric field distribution is maximum (see Figs. 2(b) and (c)). This will lead to strong nonreciprocal transmittance close to the EP, as shown in Fig. 4(a), where we calculate the upward and downward transmittance versus input intensity at the maximum transmission contrast frequency (349.4 THz). The transmittance is close to zero for both directions when the input intensity is low (less than 5 MW/cm$^2$) and the Kerr nonlinear effect is not triggered, i.e., the system operates in the linear regime. As the input intensity $I_{in}$ is increased above 5 MW/cm$^2$, there is a sharp rise in transmittance only for the downward illumination direction, as illustrated in Fig. 4(a). However, the upward illumination transmission is kept zero until the input intensity becomes approximately 10 MW/cm$^2$, which is clearly demonstrated in Fig. 4(a). When the input intensity is in the window between 5 MW/cm$^2$ to 10 MW/cm$^2$, there is a very strong and almost unitary transmission contrast triggered by self-induced nonreciprocity.

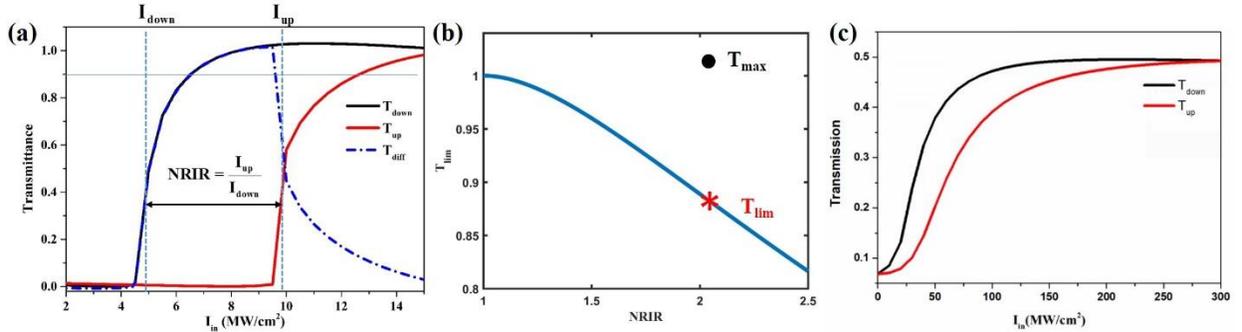

**Figure 4.** (a) Computed nonreciprocal transmittance from opposite direction illumination versus the input intensity. The dashed-dotted line depicts the absolute value of the difference between upward and downward transmittances. (b) Theoretically derived nonreciprocal transmittance limit ($T_{lim}$) versus the nonreciprocal intensity range (NRIR) (blue line) for the passive asymmetric metasurface. The maximum nonreciprocal transmittance ($T_{max}$/black dot) of the current active metasurface design substantially exceeds the passive transmittance limit under the same NRIR value. (c) Transmittance from opposite directions versus input intensity for an optimized passive metasurface based on the structure shown in Fig. 1 without utilizing the active part of the dielectric material.

The absolute value of the difference between upward and downward transmittances is also plotted with a blue dashed-dotted line in Fig. 4(a). When the input intensity is 8 MW/cm$^2$, which, interestingly, is an extremely low value compared to all previous works [21,49], the transmittance difference reaches almost unity values (0.996), where $T_{down} = 0.998$ and $T_{up} = 0.002$. Hence, the nonreciprocal response obtained here is more pronounced compared to other works [18,21,85], as it is clearly demonstrated in Table 1 that provides a direct comparison between different relevant self-induced free-standing nonlinear nonreciprocal structures. Figure 4(a) shows that ≥90% transmittance difference is achieved in a broad range of input intensities from 6.5 to 9.6 MW/cm². In order to further demonstrate and more accurately quantify the level of the nonreciprocal effect, we define the nonreciprocal intensity range (NRIR) [53], as the ratio of input intensities from opposite directions that lead to nonreciprocal transmission, calculated by the formula: NRIR = $I_{up}$ / $I_{down}$. Large NRIR values typically mean a broad range of input intensities over which large nonreciprocal transmission contrast is achieved. The proposed design can realize a relatively high NRIR value, that is equal to 2.04 ($I_{down}$ = 4.8 MW/cm² and $I_{up}$ = 9.8 MW/cm²), considering the low input intensities used to excite nonreciprocity compared to other works shown in Table 1. For the convenience of comparison, the NRIR value 2.04 can also be expressed as 3.1 dB, calculated by 10×$\log_{10}$(NRIR).

**Table 1**. Comparison between different works that demonstrate nonlinear self-induced nonreciprocal response.

| work | Bandwidth/ Center Frequency | Kerr Nonlinear Coefficient ($\chi^{(3)}$ (m²/V²)) | Thickness/ Wavelength | Signal Intensity | Isolation (or Transmission Contrast) at Best Insertion Loss |
|---|---|---|---|---|---|
| [85] | NA | 2.8×10⁻¹⁸ | 0.1 [um] / 1.5 [um] | 5 kW/cm² | -17 dB at -1.2 dB over 4.77 dB* |
| [18] | | | (2.7-6.15) [um] / 1.53 [um] | (1.5-2) MW/ cm² | Isolation of (-25.4/-35.7/-15.2) dB at insertion loss of (-0.46/-0.41/-0.044) dB over NRIR of (2.79/1.5/1.52) dB** |
| [21] | 0.6 THz / 192THz | | (1.33-5.334) [um] / 1.56 [um] | (16.8±0.001) GW/cm² | -56 dB at -0.04dB -65dB at -0.2 dB |
| This work | 2.1 THz / 349.4 THz | 6×10⁻²⁰ | 72 [nm] / 850 [nm] | 8 MW/cm² | Isolation of -26.98 dB at insertion loss of -0.017 dB over NRIR of 3.1 dB |

*Results obtained from Fig. 3(c) in [85]. **Results obtained from Figs. 3, 4, and 5 in [18].

Interestingly, it has been reported that a fundamental trade-off exists between NRIR and the maximum transmittance $T_{max}$ in self-induced nonreciprocal systems that contain nonlinear Fano resonators. This trade-off relation can be described by: $T_{max} \leq T_{lim} = 4$ NRIR/(1+NRIR)² [19,32,53], which demonstrates that maximum transmittance can be achieved only at the expense of reduced NRIR. This transmittance bound limitation results from time-reversal symmetry considerations applicable to lossless passive configurations, but the currently presented metasurface is lossy and active. This trade-off can be overcome by combining multiple nonlinear resonators at the expense of a less compact design [18,19]. As an example, by cascading two pairs of Fano and Lorentz nonlinear resonators, maximum transmittance 0.99 was obtained in [18], breaking the transmittance limit of 0.97 for this configuration. Interestingly, in the current metasurface design, the maximum nonreciprocal transmittance $T_{max}$ is close to the ideal value of one, as shown in Fig. 4(b), substantially exceeding the transmittance limit derived for a passive design characterized by the same NRIR value that is equal to $T_{lim} = 0.883$. This is because the current nonreciprocal structure has an active response and, as a result, breaks the tradeoff relation between transmission and NRIR. Without the active component, it is impossible to have almost unitary transmittance mainly due to plasmonic losses. This is demonstrated in Fig. 4(c), where the nonreciprocal transmittance can reach values close to 0.5 with much lower nonreciprocal transmission contrast when using a passive plasmonic design with zero gain in the dielectric spacer layer $(\delta = 0)$ by using the same dimensions and materials with the active configuration shown in Fig. 1.

Figure 5 demonstrates several contour plots of the computed downward and upward transmittances and their differences as a function of input intensity and frequency, when the same metasurface geometrical parameters are used compared to Fig. 2. The selected input intensity range is between 0.01 to 40 MW/cm$^2$. Both upward and downward transmittances exhibit an abrupt jump from zero to approximately one under different input intensities in a specific frequency range, depicted by the green triangle symbols in Figs. 5(a) and (b). Note that both transmittances can slightly exceed one at some input intensity values due to the active gain material used in the proposed structure. The difference between the two transmittances is plotted in Fig. 5(c) as a function of the input intensity and operating frequency values, where it is shown that the nonlinear Kerr effect is not triggered $(T_{diff} \rightarrow 0)$ when the system operates in the linear regime in the case of low input intensities ($< 5$ MW/cm$^2$). Nonlinear effects start to be triggered at higher input intensities ($> 5$ MW/cm$^2$) leading to nonreciprocity. Interestingly, the system's nonreciprocal unitary transmission contrast window depicted in Fig. 5(c) can be tuned and redshifted to lower frequencies as the input intensity is increased. Moreover, the obtained nonreciprocal unitary transmission contrast window has a relatively large bandwidth, as shown in the inset of Fig. 5(c). The unitary transmission contrast bandwidth is the region where the presented device is envisioned to operate as an almost perfect nonreciprocal optical transmission filter.

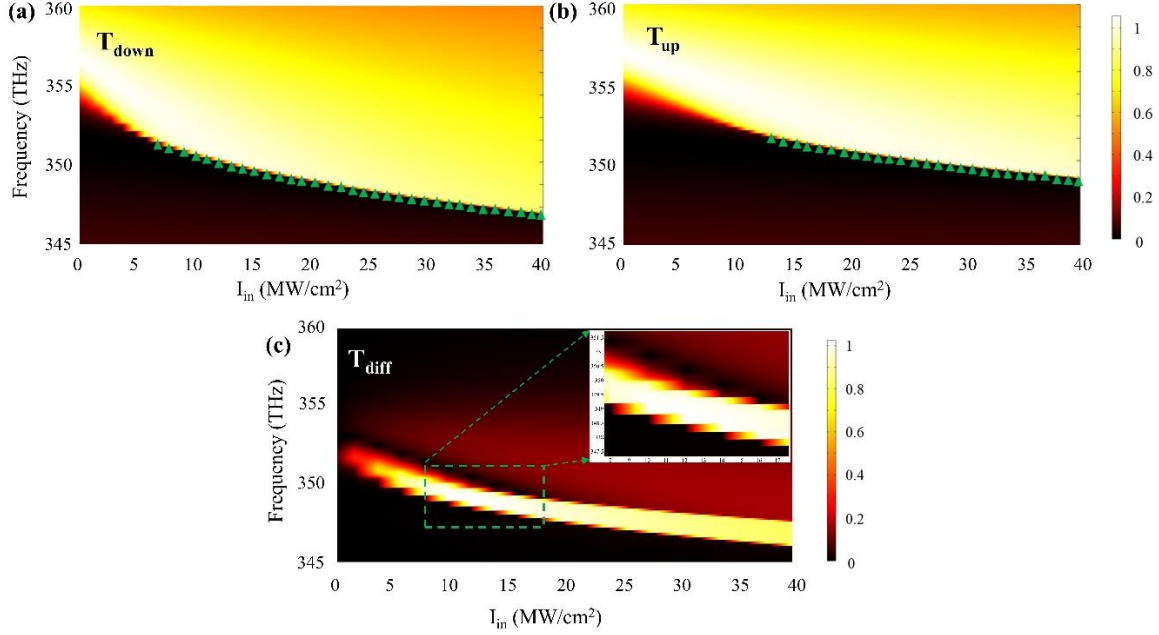

**Figure 5.** (a) Downward and (b) upward transmittances and (c) their difference as a function of input intensity and frequency for the proposed bifacial plasmonic metasurface design. The triangle symbols in (a) and (b) illustrate that there is an abrupt jump of transmittance due to the triggering of the nonlinear Kerr effect.

It should be stressed that there always exists a shortcoming in Kerr nonlinear nonreciprocal devices due to the dynamic reciprocity problem [31], which occurs when small optical signals or noise are considered in superposition with the incident large signals coming from the opposite direction. In order to investigate this so-called "dynamic reciprocity" operation regime, we simultaneously excite the presented asymmetric nonlinear structure from both directions with a relevant schematic depicted in the inset of Fig. 6. It was demonstrated in Fig. 4(a) that the maximum transmission contrast can reach almost unity values (0.996) under downward input intensity of $I_{in1} = 8$ MW/cm² at the 349.4 THz resonance frequency. Hence, we keep $I_{in1}$ fixed at 8 MW/cm², while varying the upward input intensity $I_{in2}$, and compute the output power from both sides that is calculated by $\int_C \vec{S} \cdot \vec{n}$, where the Poynting vector $\vec{S}$ crossing the upper boundary curve C is multiplied by the normal to the boundary vector $\vec{n}$ [86]. The input power is obtained by the formula: $P_{in} = I_{in} \times p \times 1[m]$, where $p$ is the period, and we assume that the y-direction arbitrary length of the structure is 1 $m$ resulting to Watt power units. Note that we operate in the frequency domain and not in time domain by using finite pulses as in ref. [31]. Thus, the power we observe at the bottom (top) is not only the transmitted power in the case of the downward (upward) incident wave, but the total power that includes the input intensity from each side. This method has been used before in [87] and our previous work to explore the different application of coherent perfect absorption [88,89]. We further calculate the output power ratio as a function of the upward input intensity $I_{in2}$ by computing the ratio of the transmitted power $P_{out1}$ or $P_{out2}$ normalized to the total input power ($P_{in1} + P_{in2}$). The result is shown in Fig. 6, where the computed transmitted power at top (port 1) is very low under small intensity $I_{in2}$, while the transmitted power at bottom (port 2) is constantly high for a range of low $I_{in2}$ values. In this case, the large nonreciprocal response (high transmission contrast between top and bottom output powers) remains unaffected. The output power at top port 1 starts to increase when $I_{in2}$ reaches the threshold value of 0.1 MW/cm². The normalized output power ratio for both ports (top and bottom) become identical when the $I_{in2}$ intensity value becomes 2.4 MW/cm² due to the dynamic reciprocity limitation. However, it is interesting that the proposed nanodevice can still achieve strong nonreciprocity at a relative broad input intensity range [17]. This region can be used to realize efficient optically controlled transmission switching devices [90]. By exciting the structure from opposite sides, reconfigurable nonreciprocity is achieved when the input intensity $I_{in2}$ is increased. This approach is self-induced, stemming from the incident illumination itself, and is different compared to other relevant works that use external magnetic field to tune the nonreciprocal response [91].

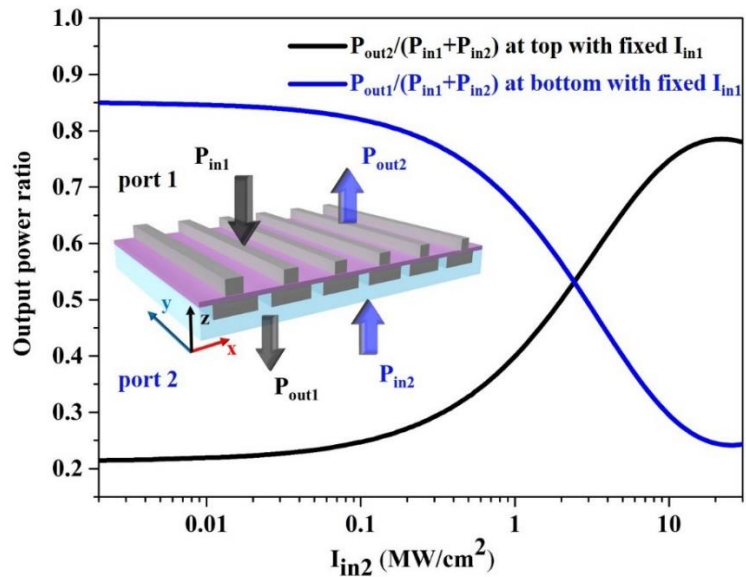

**Figure 6.** The calculated output power ratio at each side as a function of the additional excitation with varying intensity $I_{in2}$. The inset shows the metasurface schematic when simultaneously excited by two input signals from opposite directions.

As we have demonstrated before in Figs. 2(b) and (c), the electric field enhancement distribution is substantially higher for the downward illumination (port 1 excitation in the inset of Fig. 7) than it is for the upward illumination. This will definitely result in a drastic difference in the nonlinear optical response of the metasurface, not only in the fundamental frequency due to Kerr effect, as it was shown by the previous results, but also in higher harmonics generated by the nanostructure. Hence, we explore the third harmonic generation (THG) of the proposed asymmetric plasmonic metasurface structure, which is one of the most-common third-order nonlinear processes. In the THG process, an incident wave (ω) interacts with the system to produce a wave with three times the incident wave frequency (3ω) [92,93]. We measure the output power of the third-harmonic signal in the transmission direction for both excitation scenarios (up and down) by introducing an additional electromagnetic wave solver in COMSOL that couples to the fundamental frequency solver to accurately compute the THG radiation [86,93]. Next, we calculate the THG conversion efficiency which is defined as the ratio of the measured THG output power $P_{out}$ to the input power $P_{in}$. Here, the input intensity we use to excite the structure is fixed to 8 MW/cm$^2$, i.e., same with the previous computations. As expected, a striking difference in the THG conversion efficiency is observed that is clearly shown in Fig. 7. Interestingly, substantially high THG efficiency values are obtained by using a relatively low input intensity value. Thus, the THG efficiency also strongly depends on the asymmetric field enhancement within the dielectric spacer, as we have emphasized before. It is worth noting that the THG conversion efficiency for the downward illumination is increased ten times compared to the upward excitation. This result can lead to asymmetric imaging [94] and is mainly due to the asymmetric metasurface geometry. We also illustrate the THG conversion efficiency by modulating the illumination intensity and the incident frequency (see more details in the Supplementary Material [79]). The proposed design is advantageous in terms of asymmetric nonlinear light generation due to the moderate excitation intensity requirement and the ultrathin thickness of the whole structure. Hence, it is proven that nonlinear light-matter interaction at the nanoscale is a promising pathway to realize asymmetric light control [95].

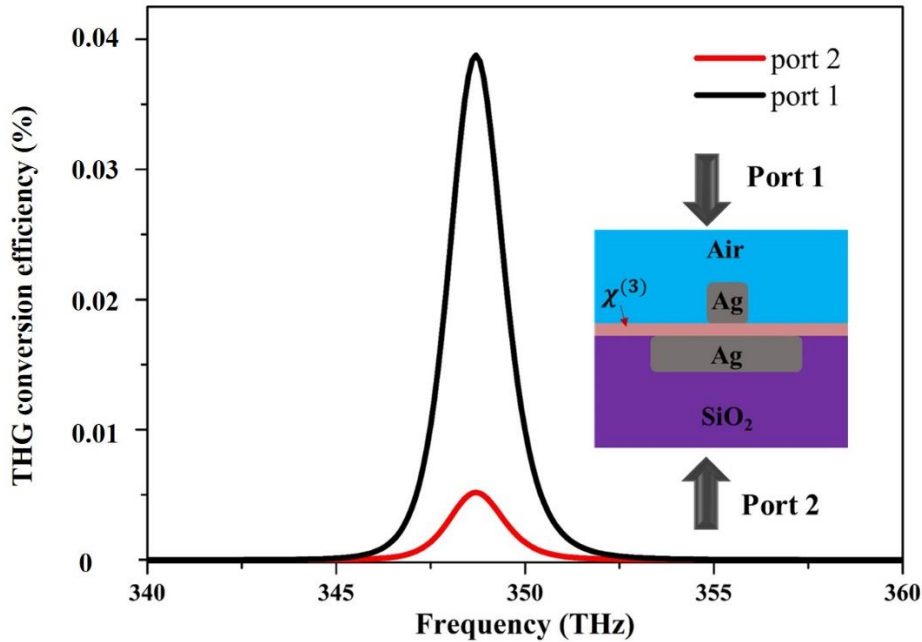

**Figure 7.** Different THG conversion efficiency obtained under opposite illumination directions. Inset: Schematic of the presented nonreciprocal metasurface structure.

## IV.  Conclusions

To conclude, we presented a nonlinear bifacial plasmonic metasurface design that achieves strong magnet-less self-induced nonreciprocal transmission at the nanoscale. The obtained strong asymmetric transmission of light is achieved close to an EP resulted from the linear PT-symmetric properties of the proposed active nanoscale system. The derived efficient nonreciprocal response can be tuned in different frequency ranges by modulating the incident laser power. Moreover, the proposed nanostructure can achieve strong nonreciprocal response for a relatively broad frequency bandwidth outperforming various previous related works mainly based on bulk photonic designs. The fundamental trade-off relation between NRIR and the maximum transmittance is also broken by the current plasmonic design due to the active material that compensates the plasmonic losses. Furthermore, the dynamic reciprocity limitation of nonlinear nonreciprocal systems is investigated and is proven that nonreciprocal response can still be achieved for a relatively broad input intensity range of the additional excitation signal. Finally, the generated third harmonic intensity under opposite direction illumination is found to be strongly unbalanced. The asymmetric response generated by the proposed self-induced plasmonic nonreciprocal system can be used in a plethora of free-standing device applications, including transmission filters to protect sensitive laser or other photonic equipment, optical power limiters, circulators, and asymmetric all-optical switches or imaging components. In a more general context, the current work proves that nonlinear light-matter interaction is an ideal pathway to realize asymmetric light control at the nanoscale.

## Acknowledgements

Christos Argyropoulos acknowledges partial support from the National Science Foundation/EPSCoR RII Track-1: Emergent Quantum Materials and Technologies (EQUATE) under (Grant No. OIA-2044049) and the Office of Naval Research Young Investigator Program (ONRYIP) (Grant No. N00014-19-1-2384). Tianjing Guo acknowledges support from the Natural National Science Foundation of China (NSFC) (12104203).

# Supplementary Material

# Nonreciprocal Transmission of Electromagnetic Waves with Nonlinear Active Plasmonic Metasurfaces

Tianjing Guo[1,#] and Christos Argyropoulos[2,*]

[1]Institute of Space Science and Technology, Nanchang University, Nanchang 330031, China

[2]Department of Electrical and Computer Engineering, University of Nebraska-Lincoln, Lincoln, Nebraska 68588, USA

[#]tianjing@ncu.edu.cn

[*]christos.argyropoulos@unl.edu

## 1. Effect of silver's nonlinearity on the transmittance of the structure

The linear permittivity of silver follows the Drude model in our simulations: $\varepsilon_{Ag} = \varepsilon_{\infty} - f_p^2 / f(f - i\gamma)$, where $f_p = 2175THz$, $\gamma = 4.35THz$, and $\varepsilon_{\infty} = 5$ [S1,S2]. Silver has a third-order nonlinear susceptibility value equal to: $\chi_{Ag}^{(3)} = 2.8 \times 10^{-19} m^2 V^{-2}$ [S3] at the currently used near-infrared (IR) frequencies. However, the fields are weakly penetrating and interacting with silver at this frequency range. As a result, its nonlinearity has little effect on the total nonlinear response of the proposed plasmonic metasurface design. In order to verify this hypothesis, we compare the upward and downward computed transmittances of the nonlinear structure with and without the silver nonlinearity when the input intensity is fixed to 8 MW/cm². The result is shown in Fig. S1. It can be clearly seen that there is no difference at the resonant frequency and total transmission response. This is depicted by the solid and dashed lines in Fig. S1 that represent the nonlinear transmission results without and with silver nonlinearity, respectively. In addition, we observe that the peak transmittance values are the same for both cases. Hence, there is no need to consider the silver's nonlinearity in the calculations performed in the main manuscript.

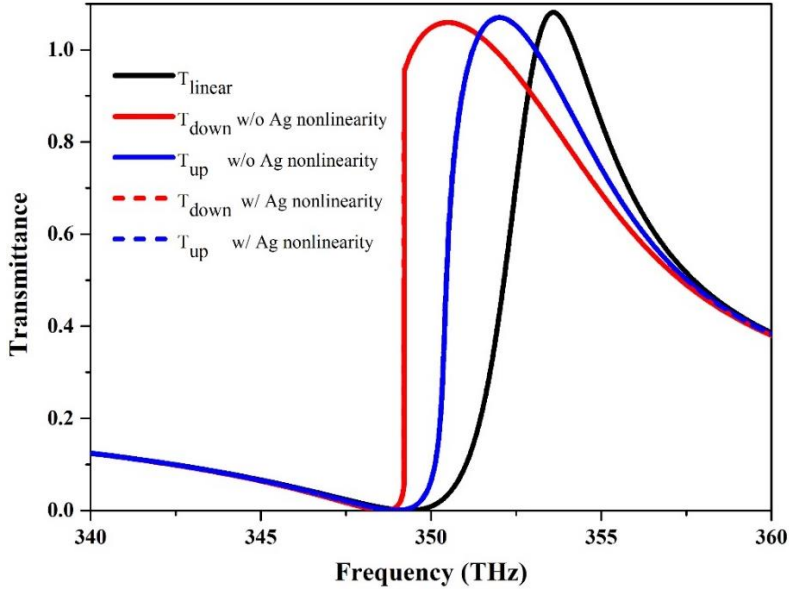

Figure S1. Linear and nonlinear transmission response under upward and downward illuminations. Solid blue and red lines represent the upward and downward transmittances without considering silver's nonlinearity. Dashed blue and red lines represent the upward and downward transmittances with silver's nonlinearity. The linear transmittance is also depicted by the black solid line.

## 2. Asymmetric harmonic generation of the nonlinear metasurface

To further explore the asymmetric harmonic generation of the nonlinear metasurface, we calculate the third-harmonic generation (THG) conversion efficiency as a function of the input intensity and incident frequency for the two excitation scenarios, i.e., downward and upward illumination. The results are shown in Figs. S2(a) and (b), respectively. The selected input intensity range is chosen to be relatively low varying between 0.01 to 40 MW/cm$^2$. We observe that the THG conversion efficiency is much higher for downward illumination than that for the upward illumination, as it can be clearly derived by inspecting the colorbars in Fig. S2(a) and (b). This can be explained due to asymmetrically enhanced field distribution within the dielectric spacer layer in the nonlinear regime under upward and downward illumination combined with the exceptional point (EP) formation dynamics. Hence, not only the nonlinear transmittance, but also the THG conversion efficiency depends on the field enhancement of the nonlinear plasmonic metasurface. It is found that the THG conversion efficiency is increased by 10 times in the case of downward illumination compared to the upward excitation scenario, spanning the entire computed frequency range. This result illustrates the advantage of the proposed design in asymmetric nonlinear light generation. Interestingly, the used moderate input intensity values lead to substantially high THG conversion efficiencies that are obtained from an extremely thin and compact metasurface design.

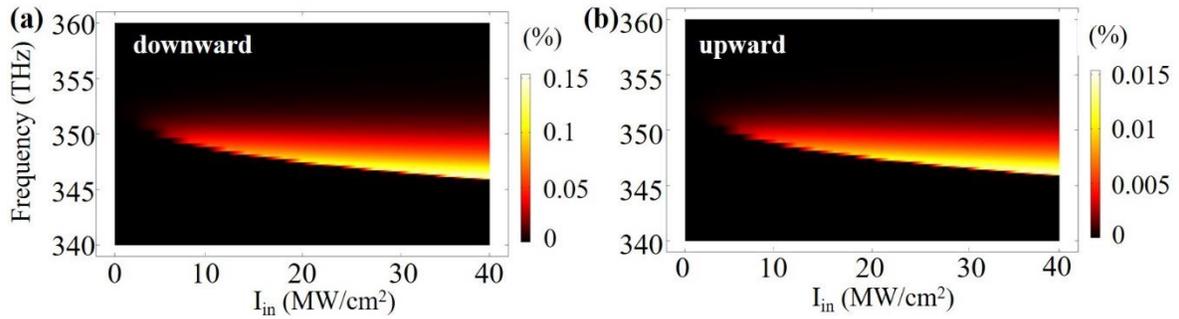

Figure S2. The proposed plasmonic metasurface design demonstrates asymmetric nonlinear harmonic generation depicted by calculating the THG conversion efficiency for (a) downward and (b) upward excitation.